\documentclass{PoS}

\usepackage{listings}
\usepackage{color}
\definecolor{gray}{rgb}{0.4,0.4,0.4}
\definecolor{darkblue}{rgb}{0.0,0.0,0.6}
\definecolor{cyan}{rgb}{0.0,0.6,0.6}
\title{Lattice QCD package GWU-code and QUDA with HIP}

\ShortTitle{latticeQCD with HIP}

\author{\speaker{Yu-Jiang Bi}\thanks{This work is supported in part by the Strategic Priority Research Program of Chinese Academy of Sciences (Grant No. XDC01040100) and the National Key Research and Development Program of China (No. 2017YFB0203200). M. Gong is supported in part by the National Science Foundation of China (NSFC) under the project No. 11775229, P. Sun is supported by Natural Science Foundation of China under grant No. 11975127, Y. Yang is supported in part by Chinese Academy of Science CAS Pioneer Hundred Talents Program.}\\
        Institute of High energy Physics,  Chinese Academy of Sciences, Beijing 100049, China\\
        }
        
\author{Yi  Xiao\\
        The Institute of Computing Technology, Chinese Academy of Sciences, Beijing 100190, China\\
        E-mail: \email{louisxcode@gmail.com}}

\author{Ming Gong\\
        Institute of High energy Physics,  Chinese Academy of Sciences, Beijing 100049, China}

\author{Wei-Yi Guo\\
        Department of Physics, University of Warwick, Coventry, CV4 7AL, United Kingdom}

\author{Peng Sun\\
        Nanjing Normal University, Nanjing, Jiangsu, 210023, China\\
        E-mail: \email{06260@njnu.edu.cn}}

\author{Shun Xu\\
        Computer Network Information Center, Chinese Academy of Sciences, Beijing 100190, China}

\author{Yi-Bo Yang\\
        CAS Key Laboratory of Theoretical Physics, Institute of Theoretical Physics, Chinese Academy of Sciences, Beijing 100190, China\\
        E-mail: \email{ybyang@itp.ac.cn}}

\abstract{The open source HIP platform for GPU computing provides an uniform framework to support both the NVIDIA and AMD GPUs, and also the possibility to porting the CUDA code to the HIP-compatible one. We present the porting progress on the Overlap fermion inverter (GWU-code) and also the general Lattice QCD inverter package - QUDA. The manual of using QUDA on HIP and also the tips of porting general CUDA code into the HIP framework are also provided.}

\FullConference{37th International Symposium on Lattice Field Theory - Lattice2019\\
		16-22 June 2019\\
		Wuhan, China}

\begin{document}

\section{Introduction and background}

The engineering and energy efficiency constraints push the modern supercomputer architecture to the multi-level parallelism, and heterogeneous computing architectures such as CPU+GPU are widely used in top500 supercomputers, including the most recent fastest two, Summit and Sierra. Most of the performance on such a computer come from the Nvidia GPU V100, and an efficient code is essential to benefit the related Lattice QCD calculation from those machines. There are already quite a few package can support the Nvidia GPU code platform CUDA with good performance and also multi-GPU scaling,  including QUDA (for most of the fermion actions)~\cite{Clark:2009wm,Babich:2011np,Clark:2016rdz},  GRID (for the domain wall fermion and etc.)~\cite{Boyle:2015tjk}, GWU-code (for the overlap and clover fermion)~\cite{Alexandru:2011ee,Alexandru:2011sc}, and so on.

On the other hand, the efficient code on the AMD GPU falls behind except some effects with OpenCL (e.g., CL2QCD~\cite{Philipsen:2014mra}), while the peak performance of the AMD GPU have caught up and the E-flops supercomputer ``Frontier" with AMD GPU will be built in US by 2021. In recent years, an open source HIP platform is promoted to support both the Nvidia and AMD GPUs, which also provide the possibility to porting the CUDA codes to the HIP platform. Based on this, writing the code from scratch and implementing hundreds features needed by the lattice QCD calculation, would be avoid by porting the existed CUDA codes. In this proceeding, we will present our finding on using the package GWU-code and QUDA on the AMD GPU through the HIP platform, with a summary on the known issues.

\section{Porting tips and compiling manual of QUDA}

Generally, the porting can be separated into 3 stages, convert the code with hipify-perl, patch the codes manually to satisfy the requirement of compiler, replace the unsupported features to avoid the runtime crash. Let us taking the porting of QUDA as example:

1).  Convert the code with hipify-perl. The hipify-perl is a perl script to map the name of the CUDA functions to that of their HIP counterpart, and also the CUDA head files. If the script meets some unknown words starting with "cu", message ``warning:... : unsupported device function" will be thrown out, while the conversion will continue.  

\begin{figure}
  \centering
  \includegraphics[width=0.6\textwidth]{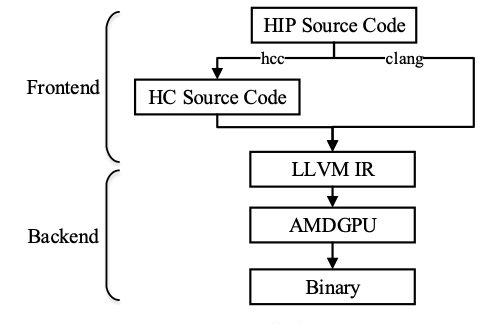}
  \caption{HIPCC Compilation Process. The clang compiler skips the step to generate the HC source code.}
  \label{fig:hipcc}
  %\label{fig:ZAV}
\end{figure}

2).  Patch the codes manually to satisfy the requirement of compiler. Currently, QUDA is compiled by hip-clang, instead of HCC (Heterogeneous Compute Compiler). HCC is HC compiler, which is C ++ AMP syntax language with HSA Extend~\cite{hsa_extend}. HCC will translate HIP kernel syntax into C ++ AMP syntax by using functional or macro grid launch, while certain QUDA device function used complicated class and template will cause syntax or runtime error. In the other hand, hip-clang is a hip kernel syntax supported LLVM frontend as shown in Fig.~\ref{fig:hipcc}. By setting the environment variable HIP\_PLATFORM to clang, hip-clang will take over the compiling and compile CUDA-like syntax source code to LLVM IR directly~\cite{GPGPU}, and then the AMD GPU backend of LLVM will compile the LLVM IR to binary. 
The patches we applied include:

2.1) Set CMAKE\_CXX\_SYSROOT\_FLAG\_CODE to add the .cu suffix to the \\
CMAKE\_CXX\_SOURCE\_FILE\_EXTENSIONS, and then use hip-clang to compile both the .cpp and .cu files. Note that the flag ``-g" should be avoid and ``-\_STRICT\_ASNI\_ -O3" is necessary to make the code works well. Compile clover\_deriv\_quda.cu and gauge\_stout.cu will crash the compiler and add the flag "-fno-inline" would be a choice to avoid it.

2.2) Fix the functions with different prototype, likes hipGetErrorString, hipGetErrorName, hipPointerGetAttributes, hipMemcpyHtoDAsync and etc. At the same time, some of the functions likes blasCgetrfBatched and blasCgetriBatched are replaced with the CPU version.

2.3) Add the \_\_host\_\_ flags in front of the  \_\_device \_\_ functions which used in the CPU code. 

2.4) Rewrite the ptx code into the normal C++ codes.

2.5) Move the declare of the shared memory into the body of the functions, as it can not be located in the link stage.

3). Replace the unsupported features to avoid the runtime crash. The major changes include:

3.2) Suppose the memoryType is host if hipPointerGetAttributes return an error, and comment out checkCudaError() in the constructors;

3.3) Limited the maximum threads used in the tuning to 512 or even smaller number, if certain function crashes on the AMD GPU with more threads.

3.4) Use the global function to copy the constant array needed by multi\_blas\_kernel to the global memory, as the function hipmemcpyToSymbol doesn't work correctly:
 \begin{lstlisting}
 __global__ void set_Amatix(signed char *ref) {
 	int idx = blockIdx.x * blockDim.x + threadIdx.x;
	if(idx>=MAX_MATRIX_SIZE)
		return;
	Amatrix_d[idx]=ref[idx];
}
  \end{lstlisting}
 \begin{lstlisting}
 	signed char *A_d;
 	hipMalloc(&A_d, MAX_MATRIX_SIZE);
	hipMemcpy(A_d,A,MAX_MATRIX_SIZE,hipMemcpyHostToDevice);
 	set_Amatix<<<256,MAX_MATRIX_SIZE/256>>>(A_d);
	hipDeviceSynchronize();hipFree(A_d);
 \end{lstlisting}
 
3.5) In the function multiblas, multireduce and ComputeVUV, pass the parameters though the argument class of the global function likes 
 \begin{lstlisting}
__global__ func(Arg arg){...}
 \end{lstlisting}
can make the performance to be extremely low. Such a problem can be avoid by copying the argument class to the GPU memory first, and use the its reference as the argument:

 \begin{lstlisting}
__global__ func(Arg &arg){...}
 \end{lstlisting}
 \begin{lstlisting}
	Arg *arg_d;
	hipMalloc(&arg_d, sizeof(Arg));
	hipMemcpy(arg_d,&arg,sizeof(Arg),hipMemcpyHostToDevice);
	func(*arg_d);
	hipDeviceSynchronize();
	hipFree(arg_d);
 \end{lstlisting}
 
 Other minor hacks can be found in the present branch on the github:
 \begin{lstlisting}
	https://github.com/lattice/quda/tree/rocm-devel.
 \end{lstlisting}
 
 \begin{table}
  \centering{
  \begin{tabular}{l | c | c  }
    \hline
     Module name & Ported & Tested \\
    \hline
   QUDA\_DIRAC\_WILSON     & yes & yes  \\
    \hline
    QUDA\_DIRAC\_CLOVER    &  yes & yes  \\
    \hline
    QUDA\_CONTRACT   &  yes & yes  \\
    \hline
    QUDA\_COVDEV    &  yes & yes  \\
    \hline
    QUDA\_DIRAC\_STAGGERED    &  yes & no  \\
    \hline
    QUDA\_FORCE\_GAUGE    &  yes & no  \\
    \hline
    QUDA\_DIRAC\_DOMAIN\_WALL    &  no & no  \\
    \hline
    QUDA\_DIRAC\_TWISTED\_MASS    &  no & no  \\
    \hline
    QUDA\_DIRAC\_TWISTED\_CLOVER    &  no & no  \\
    \hline
    QUDA\_DIRAC\_CLOVER\_HASENBUSCH    &  no & no  \\
    \hline
    QUDA\_DIRAC\_NDEG\_TWISTED\_MASS    &  no & no  \\
    \hline
    QUDA\_LINK\_ASQTAD    &  no & no  \\
    \hline
    QUDA\_LINK\_HISQ    &  no & no  \\
    \hline
    QUDA\_FORCE\_HISQ   &  no & no  \\
    \hline
    QUDA\_GAUGE\_TOOLS    &  no & no  \\
    \hline
    QUDA\_GAUGE\_ALG   &  no & no  \\
    \hline
    QUDA\_DYNAMIC\_CLOVER   &  no & no  \\
    \hline
  \end{tabular}
  \caption{Summary of the porting progress in term of QUDA build options. All the parts needed for a Multigrid inverter of the Clover fermion have been done.}
  \label{tab:port}
  }
\end{table}
 
The present code merged the recent released QUDA 1.0.0. In term of the QUDA build options, the porting progress are listed in Table~\ref{tab:port}.
  
 \section{GWU-code case}
 
 Comparing to QUDA, porting GWU-code is much simpler. GWU-code use the macro to generate the D-slash GPU kernel without any device functions, and implement the vector operator on GPU with the CUDA Thrust~\cite{Alexandru:2011ee,Alexandru:2011sc}. Thus one just need to replace the Thrust library with the rocthrust after the code has been converted with hipify-perl. The function thrust::reduce can be very slow with double precision in certain version, but it can be replaced by the other functions with normal performance.

\section{Performance and issues}

Our test is based on AMD MI60 GPU and Nvidia V100 GPU. The D-slash performance is summarized in Table~\ref{tab:dslash}. The QUDA dslash performance is somehow lower as this kernel is much more complicated and then can not be fully optimized with hip-clang at present. 

\begin{table}
  \centering{
  \begin{tabular}{c | c | c | c | c }
    \hline
        & Peak FP32 (TFlops) & Bandwidth (TB) & GWU-code (TFlops) & QUDA (TFlops) \\
    \hline
   Nvidia V100     & 14.7 & 0.9  & 0.80 & 1.11  \\
    \hline
    AMD MI60    & 14 (for PCIe) & up to 1.0 & 0.54 & 0.48 \\
    \hline
  \end{tabular}
  \caption{The single precision peak performance and memory bandwidth of V100 and MI60, v.s. the D-slash performances using the GWU-code and QUDA with the same single precision. The SU(3) is suppressed to 12 real numbers in both the GWU-code and QUDA cases to save the memory bandwidth. Note that the tests on V100 uses CUDA, not HIP. }
  \label{tab:dslash}
  }
\end{table}

As the practical application of QUDA, the multigrid inverter works correctly while the performance is not very permising. With the largest $96^3\times192$ lattice we tested, the total performance with  324 MI60 GPUs is around 10 TFlops, using a 3-level multigrid layouts (4,4,4,4) and (2,2,2,2). 

In the GWU-code side, 200 pairs of the Overlap eigensystem of the HYP smeared $24^3\times 64$ RBC configuration at $a$=0.11fm can be generated with 4 MI60 GPU in 4 hours, and the similar calculation with E5-2698v3 at 2.3 GHz requires 1024 cores for 2 hours. The test with larger size is in progress.

\section{Summary}

In summary, we port two Lattice QCD CUDA packages, GWU-code and QUDA to the AMD GPU platform using HIP. The performance is around half of that on the CUDA when the memory bandwidth of them are similar. The multigrid inverter of QUDA works correctly with the lattice as large as  $96^3\times192$, and the overlap eigensystem can be generated correctly with GWU-code. We will try to optimize the performance and scaling in the further study.  All the present test using HIP are done on AMD GPUs, that on Nvidia GPUs will be also investigated. 

\bibliographystyle{unsrt}
\bibliography{bibliography.bib}

\end{document}